\documentclass{ws-ijn}

\begin{document}

\markboth{Yu. G. Arapov, S. V. Gudina, G.I. Harus, et al}
{Transport properties of 2D-electron gas in a n-InGaAs/GaAs DQW in a 
vicinity ...}

%
\catchline{}{}{}{}{}
%

\title{Transport properties of 2D-electron gas in a n-InGaAs/GaAs DQW in a 
vicinity of low magnetic-field-induced Hall insulator--quantum Hall liquid 
transition 
}

\author{\footnotesize Yu.G. Arapov, S.V. Gudina, G.I. Harus, V.N. Neverov, N.G. Shelushinina, 
M.V. Yakunin, S.M. Podgornyh }

\address{Institute of Metal Physics Ural Division RAS, \\ Ekaterinburg 620219, Russia
}

\author{E.A. Uskova, B.N. Zvonkov}

\address{Physical-Technical Institute at Nizhnii Novgorod State University,\\ 
Nizhnii Novgorod, Russia
}

\maketitle

\begin{history}
\received{(Day Month Year)}
\revised{(Day Month Year)}
\end{history}

\begin{abstract}
The resistivity ($\rho $) of low mobility dilute 2D-electron 
gas in a n-InGaAs/GaAs double quantum well (DQW) exhibits the monotonic
"insulating-like" temperature dependence ($d\rho /dT<0$) at $T=1.8-70$K in 
zero magnetic field. This temperature interval corresponds to a ballistic 
regime ($k_{B}T\tau /\hbar >0.1-3.5$) for our samples, and the electron density is 
on a "insulating" side of the so-called $B=0$ 2D metal-insulator transition.
We show that the observed localization and Landau quantization is due to the $\sigma_{xy}(T)$
anomalous $T$-dependence.

\end{abstract}

\section{Introduction}	

The coexistance of so-called two-dimensional (2D) "metallic" and "insulating" phases, 
identified in n-Si-MOSFET's by Kravchenko and Pudalov \cite{a1,{a2}}, is still a  
subject of considerable interest and controversy. While the effectively 
"metallic" and "insulating" character of phases is observed experimentally 
in many 2D- systems, the microscopic origin of the both 
phases, particularly in vicinity of the metal-insulator transition (2D- MIT) 
, is a mystery yet.

Recently there has been a great renewal of interest to the low
magnetic-field-induced Hall insulator--quantum Hall liquid (HI-QHL) 
transitions \cite{{a3},{a4},{a5},{a6}}. According to the scaling theory of localization in à 
noninteracting 2D - systems in zero magnetic field, at low temperatures only the localized states 
should exist. In the presence of à strong perpendicular 
magnetic field the Landau quantization becomes important\cite{a7}. 
The "fate" of electronic states from being extended at à strong magnetic 
field\cite{a7} to being localized at $B= 0$ was at first explained by Laughlin and 
Khmelnitskii\cite{a8}. It was argued that to be consistent with the scaling 
theory, the extended states should "float up" indefinitely in energy in the limit of $B=0$
. An alternative to this "floating-up" picture is that the 
extended states could be destroyed by increasing disorder or with decreasing
magnetic field they merge, forming a metallic state in $B=0$\cite{a13}.

To date, an interesting but unsettled issue is whether the observed direct 
transitions from an insulating state in $B=0$ to a conducting ones for high 
Landau-level filling factor ($\nu > 3$) are genuine quantum phase transitions (QPT). 
Hackestein\cite{a9} claimed that there is no any QPT. The low field  
transitions to the quantum Hall effect (QHE) states with $\nu \ge 3$ are only crossovers from quantum corrections regime to Landau quantization (LQ) after 
$\omega_c \tau > 1$. Kim et al. and Huang et al.  
made an attempt to overcome the  Huckenstein's doubt
about the QPT truth in their recent papers \cite{{a5},{a6}}.

They observed the crossover from the low-field localization to QHE 
that covers a wide range of magnetic fields. 
With increasing $B$, Shubnikov-de Haas oscillations (SHO) appear at $B_s$,  $\rho_{xx}=\rho_{xy}$
at $B_a$ and $\rho_{xx}$ becomes $T$-independent at  $B_c$ ($B_s<B_a<B_c$). Thus the observed well-defined 
critical points of $\rho_{xx}(B, T)$ don't correspond to the crossover from localization to LQ when $\rho_{xx}=\rho_{xy}$.
Experimental data in the vicinity of the critical point $B_C$ show good scaling behaviour confirming that HI--QHL transition is a genuine QPT.

On the other hand, it is argued 
that such a low-field transition is not a phase transition, but can
result from the interplay of the classical cyclotron motion and the 
electron-electron interaction quantum correction $\delta \sigma ^{ee}$ 
to the Drude conductivity in the diffusive regime ($k_{B}T\tau /\hbar <1$) \cite{a10,{a11}}.

The HI-QHL transition in low magnetic fields was investigated by different 
authors on the "insulating" side of the 2D- MIT ($B=0$) \cite{a3} or on the "metallic" 
side one, but in the diffusive regime, where resistivity have an "isulating-like" behaviour due to weak localisation or 
ee-interaction effects\cite{{a4},{a5}}. We call the phase "insulating", when $\rho_{xx}$
diverges as $T \to 0 (d\rho /dT<0$) as opposed to the quantum Hall liquid phase, where $\rho_{xy} \to h/\nu e^2$ 
and $\rho_{xx} \to 0$ with vanishing T. The boundary is then simply the $B$-field, where the $\rho_{xx}(B,T)$
dependence changes direction ($d\rho /dT>0$). 

In this report we observed the HI-QHL 
transition at the relatively high temperatures (ballistic regime) in 
the low mobility dilute 2D-electron gas confined within 
GaAs/n-InGaAs/GaAs DQW. The transport in the vicinity of à quantum phase 
transition is discussed. We have shown that the appearence of different critical points ($B_s, B_a, B_c)$\cite{a6}
is possibly caused by the anomalous temperature dependence of $\sigma_{xy}(B,T)$.

\section{Experimental set-up}

We used $n$-type modulation doped GaAs/n-InGaAs/GaAs DQW samples. Here we 
present the data obtained for one of samples with the following parameters 
at low temperatures: the electron density $n_s=2.3 \times 10^{11}$cm$^{-2}$ 
and the mobility of $1.16 \times 10^4$cm$^{2}$/Vs ($k_{F}l_{tr}=10.6$) . Four- 
terminal longitudinal ($\rho _{xx})$ and Hall ($\rho _{xy})$ resistivity 
measurements were carried out in a "Quantum Design" equipment using a 
standard low frequency (10 Hz ) lock-in technique in a wide temperature 
interval (1.8-300K) and in tilted magnetic fields up to 9.0T. A $\le 1 
\mu $A driving current through Hall bar was chosen to avoid heating 
effects. 

\section{Experimental results and discussion}

\textbf{1}. Fig.1 shows a set of resistance versus temperature 
dependences. For the first time the insulating-like behavior ($d\rho /dT<0$) 
of the resistance dependence in a whole temperature interval up to $T \cong 
$70K ( $T/T_{F} \quad  \cong 0.65$) is observed. For the material of our 
samples, this temperature interval corresponds to a ballistic regime 
($k_BT\tau /\hbar >0.1$)\cite{a12}. 
There exist a few theoretical models to explain the apparent insulating $\rho (T)$ 
behaviour of the 2D metallic state at high temperatures ($T < T_{F}$). A 
simple non-interacting picture for the apparent insulating behaviour of 
metallic 2D systems at high $T$ is the $T$-dependent scattering of a 
non-degenerate electron gas\cite{a2}. On the other hand, for the degenerate electron 
gas, the temperature-dependent screening\cite{a2}
and electron-electron interaction\cite{a12} effects play important quantative roles 
in determening the type of resistivity temperature dependence and producting the 
effective insulating-like behaviour.

\begin{figure}[t]
\psfig{file=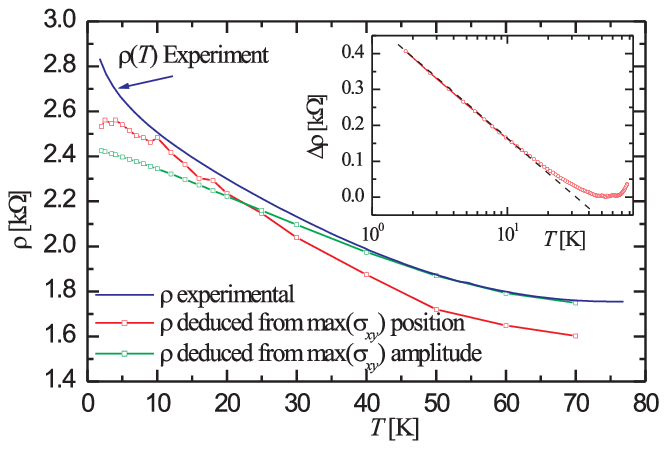,width=6cm} \psfig{file=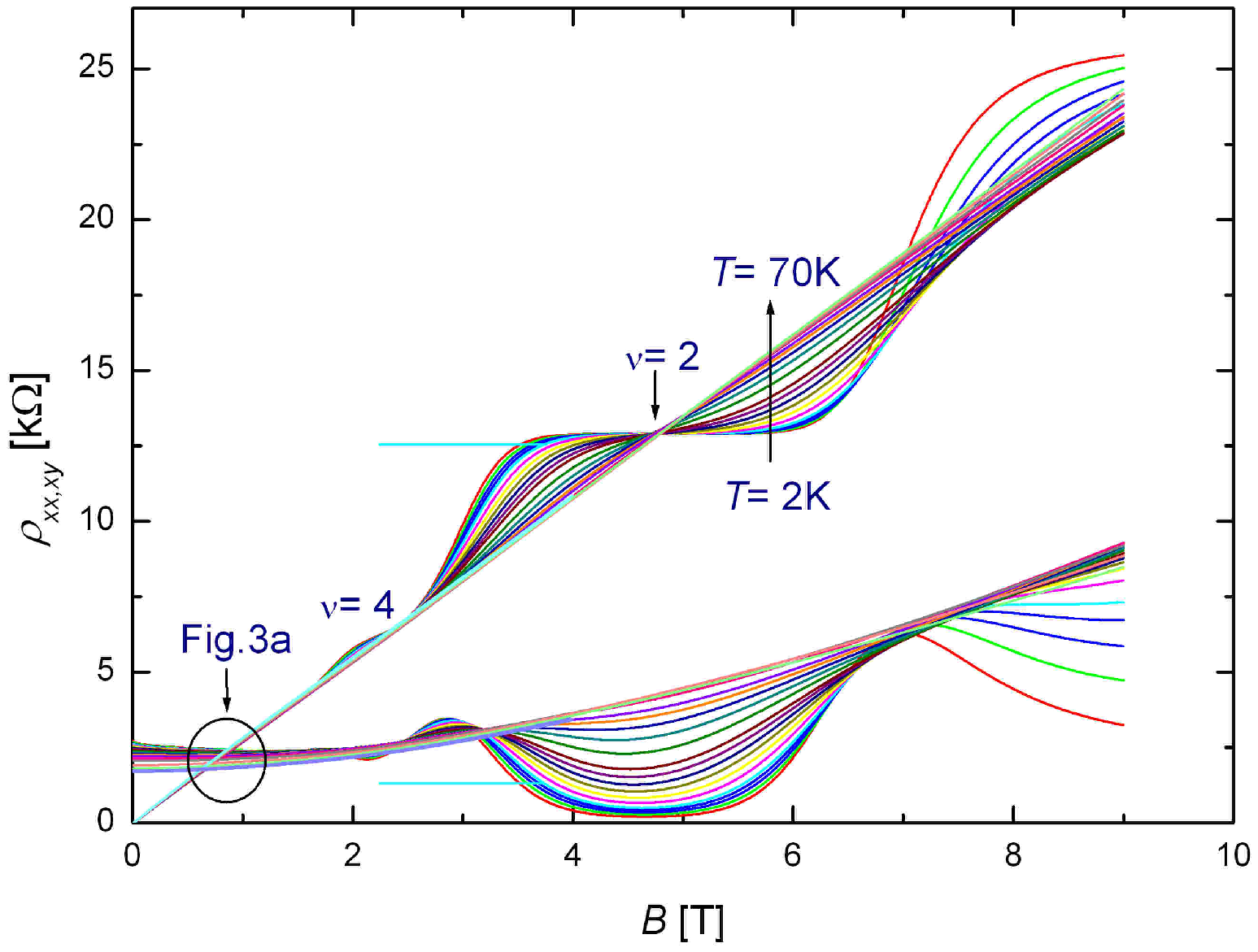,width=6cm} 
\vspace*{8pt}
\caption{Resistivity for sample 3982b at $B_\perp =0$  vs temperature. Difference between the 
experimental results and those deduced from $\sigma_{xy}(B,T)$ resistivity is shown in the inset.}
\caption{Four-terminal resistivity measurements of $\rho_{xx}(B)$ and $\rho_{xy}(B)$ at $T=2 \div 70$K.}
\end{figure}

This 
dependence could be quantitatively described by three contributions\cite{a14} 
\begin{equation}\label{eq:1}
\rho (T)=\rho _D(T) + \delta \rho _{WL}(T) +\delta \rho 
_{eei}(T), 
\end{equation}
where $\rho _{D}(T)$ is the Drude resistance, $\delta \rho _{WL}(T)$ 
and $\delta \rho _{eei}(T)$ are the weak localization and interaction 
contributions, respectively. Now, that we have experimentally determined 
$\rho _{D}(T) = \sigma _{D}^{- 1}(T)$ from analysis of $\sigma 
_{xy}(B,T)$  (see below), we are in a position to extract the bare $\delta 
\rho _{WL}(T) + \delta \rho _{WL}(T)$, according to Eq.~(\ref{eq:1}), by 
subtracting the Drude resistance from the measured zero-field resistance 
(Fig.~1). The obtained two corrections show logarithmic temperature 
dependence (see inset on Fig.1) Origin of the temperature dependences of Drude 
resistance and quantum corrections will be analysed elsewhere\cite{ad1}.

\textbf{2.} In Figs. 2-4 we plot the measurement data of longitudinal 
magnetoresistivity $\rho _{xx}(B,T)$, $\rho _{xy}(B,T)$ and magnetoconductivity $\sigma _{xx}(B,T)$, $\sigma 
_{xy}(B,T)$ over the temperature range $T = 1.8-50.0$K. Fig.2 shows the well 
expressed SHO in 
$\rho _{xx}(B,T)$ and of QHE plateaus in 
$\rho_{xy}(B,T)$. Pronounced minima in $\rho 
_{xx}(B,T)$ traces observed at filling factors $v=2,4,6$, which 
are accompanied by QHE plateaus in $\rho _{xy}(B,T)$ at 
corresponding magnetic fields, are used for evaluation of the carrier 
density. The temperature-independent magnetic field positions of SHO 
minima allow us to make an important assertions that the carrier density 
for this sample up to $T \cong 50$K is unchanged and therefore such an unusual 
temperature dependence of resistivity is not due to the carrier density 
change. In the low magnetic fields perpendicular to the 2DEG ($B_{ \bot })$ 
the negative magnetoresistance (NMR) on a $\rho _{xx}(B,T)$ was observed. 
The so-called temperature independent ($T_{ind}$) point is seen (Fig. 3a) at some 
$B_{cr}$. At temperatures $T>6.0$K this point begins to "wash out". May be 
at this value of magnetic field we have $\omega _{c}\tau =1$?\cite{a10,{a11}} 
But why doesn't this point coincide with the point where $\rho 
_{xx}(B,T)=\rho _{xy}(B,T)$ (see Fig.3a)?  We think that this lack of coincidence is connected with 
the beginning of the transition from diffusive to ballistic regime. 

\begin{figure}[b]
\centerline{\psfig{file=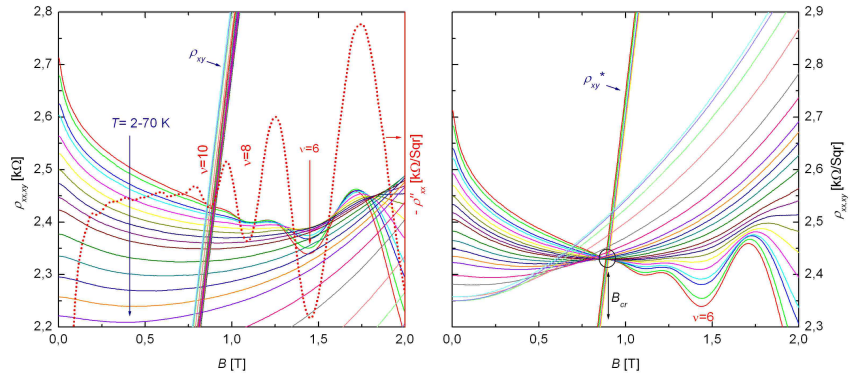,width=12cm}}
\vspace*{8pt}
\caption{(a) $\rho_{xx}(B,T)$ and $\rho_{xy}(B,T)$. There is a $T_{ind}$-point,
where $\rho_{xx}(B,T)$ traces cross itself in a $B_{cr}$. 
(b) corrected $\rho_{xx}^{*}(B,T)$ and $\rho_{xy}^{*}(B,T)$. At the critical point $B^*_{cr}$:~  
$\rho_{xx}^{*}(B^*_{cr},T)= \rho_{xy}^{*}(B^*_{cr},T)$.
}
\end{figure}

The Hall resistance in a low magnetic 
fields ($\rho _{xy}(B,T)$) is  also strongly temperature 
dependent. Before analyzing the role of the quantum corrections in the 
diffusive ($k_{B}T\tau /\hbar \ll 1$) and ballistic  ($k_{B}T\tau 
/\hbar >0.1$) regimes in a behaviour of the resistance temperature dependence in a 
$B=0$ shown in Fig.1, let us estimate the possible contribution from other 
temperature dependent factors.

Observation of SHO, QHE and an "insulator-QH transition" in the insulating regime 
of low density 2D-electrons is somewhat remarkable, since it was originally 
believed that the insulating behavior of $\rho(T)$ at $B = 0$ in this 
temperature range 
is related to the non-degenerate nature of the low density hole gas \cite{a1,a2}. Note 
that if the insulating $\rho (T)$ at $T < T_{F}$ has a classical 
origin, as in Ref.1,2, it would be unlikely for the sample to show a negative 
magnetoresistivity and the "insulator-QH transition" as in Fig.2 because the 
classical Drude magnetoresistivity is zero. At present, it is unclear if a 
more elaborated semi-classical model can reproduce the insulating-like $T$-dependent 
zero field resistivity and an "insulator- QH transition" at high $T$ at the 
same time. 

\begin{figure}[t]
\psfig{file=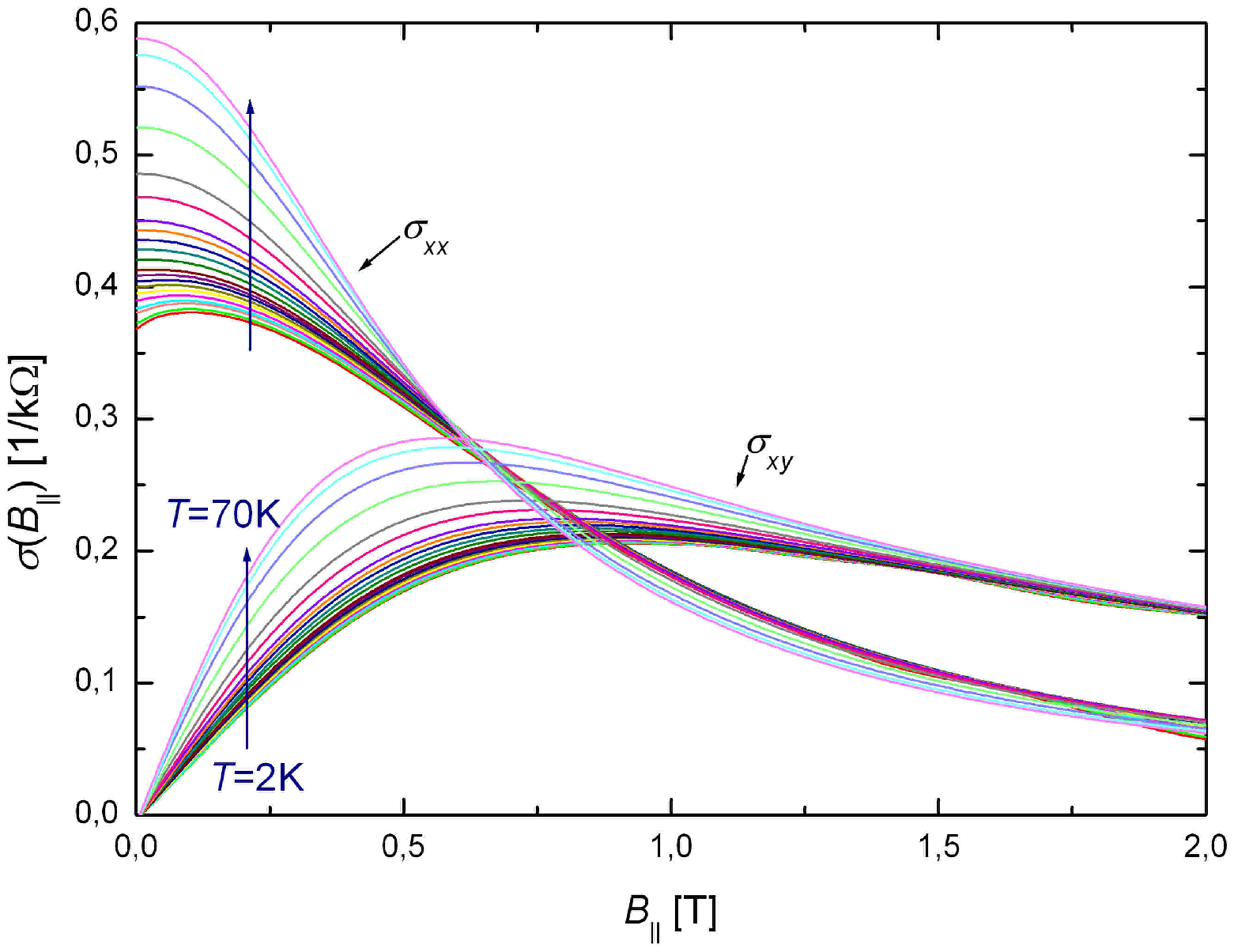,width=6cm} \psfig{file=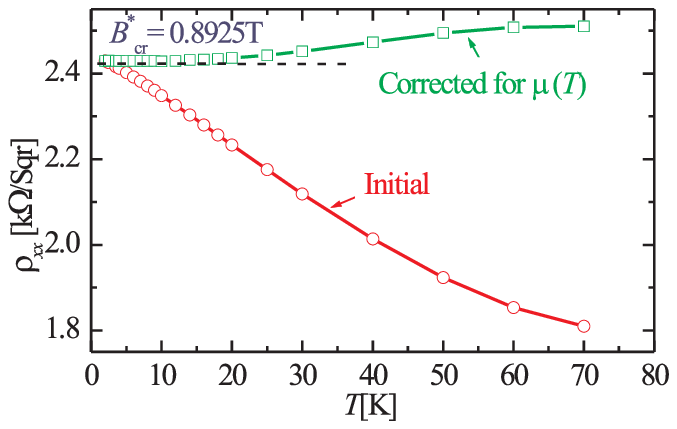,width=6cm}
\vspace*{8pt}
\caption{Converted conductivities $\sigma_{xx}(B,T)$ and $\sigma_{xy}(B,T)$
at varies temperatures $T=2\div 70$K.}
\caption{Temperature dependence of initial and corrected $\rho_{xx}(T)$ in a $B_{cr}$.}
\end{figure}

\textbf{3.} Let us pay attention to a temperature dependence of the 
magnetoconductivity $\sigma _{xy}(B,T)$ (Fig.4). It is well known that 
classical part of $\sigma _{xy}(B,T)$ should be temperature independent at 
such a degeneracy of electron gas ($T_{F}/T \cong 2-50$). All 
contradictions are resolved if we suppose that the electron mobility for our 
case is temperature dependent: $\mu (T)$ \cite{a2,ad1,{a15}}. An analysis of the 
temperature dependence of the magnetoconductivity $\sigma _{xy}(B,T)$ 
allows us to find it. According to the Drude theory, $\sigma _{xy}(B)$ 
have a maximum at $\mu B=1$ (or $\omega _{c}\tau =1$) and this value is 
equal to $\sigma _{D}/2$, where $\sigma _{D}$ is Drude conductivity. 
Doted lines in Fig. 1 represent $\sigma _{D}^{ - 1}(T)$, dedused from $\sigma _{xy}$ maximum amplitude, and $\rho 
(T)=(en_{s}\mu (T))^{-1}$, where $\mu (T)$ is obtained from $\mu 
B=1$. It is seen that these dependences are close to the experimental ones (solid line). 
However this poses a question: if the conventional Drude theory is 
applicable, what is the reason of the mobility temperature dependence? We 
attribute the $T$-dependence of the mobility and hence the Drude conductivity to 
the $T$-dependence of $\tau (T)$\cite{a17}.

Then we extracted the temperature dependence of $\sigma 
_{D}(T)$ due to $\mu (T)$ from the experimental dependence $\sigma 
_{xy}(B,T)$ and $\sigma _{xx}(B,T)$. After that we calculated the 
resistivity tensor component $\rho _{xx}^{\ast }(B,T)$  using 
the new $\sigma _{xx}^{\ast }(B,T)$ and the calculated $\sigma 
_{xy}^{\ast }(B)$ \cite{ad1,a15}. The result of this procedure is presented at 
Fig.3b. It is seen very well now that $T_{ind}$ point coincides with the 
point where $\rho _{xx}(B,T) = \rho _{xy}(B,T)$ and its "washing out" considerabbly decreases (Fig. 5).

In summary, we have presented the studies of "isulating-like" transport properties of the 
low mobility 2DEG in a n-InGaAs/CaAs DQW in a wide temperature interval $T=1.8\div 70$K
(diffusive and ballistic regimes). The presence of anomalous temperature dependence of $\sigma_{xy}(B,T)$ 
challenges the conventional understanding of low magnetic-field-induced Hall insulator -- quantum 
Hall liquid transitions.

\section*{Acknowledgments}

The work was supported in part by: Russian Foundation for Basic Research 
RFBR, grants 04-02-16614 and 05-02-16206; CRDF and Ministry of education and 
science of Russian Federation, grant Y1-P-05-14 (Ek-05 [X1]); program of 
Russian Academy of Sciences "Low-dimensional quantum heterostructures"; Ural 
division of Russian Academy of Sciences, grant for young scientists.

\end{document}